# A >99.9%-fidelity quantum-dot spin qubit with coherence limited by charge noise

J. Yoneda[1,2,*], K. Takeda[1,2], T. Otsuka[1,2,3], T. Nakajima[1,2], M. R. Delbecq[1,2], G. Allison[1], T. Honda[4], T. Kodera[4], S. Oda[4], Y. Hoshi[5], N. Usami[6], K. M. Itoh[7], S. Tarucha[1,2]

[1] RIKEN Center for Emergent Matter Science, RIKEN, 2-1 Hirosawa, Wako, Saitama 351-0198, Japan

[2] Department of Applied Physics, University of Tokyo, Tokyo 113-8656, Japan

[3] JST, PRESTO, 4-1-8 Honcho, Kawaguchi, Saitama, 332-0012, Japan

[4] Department of Physical Electronics, Tokyo Institute of Technology, Meguro-ku, Tokyo 152-8552, Japan

[5] Institute of Industrial Science, University of Tokyo, Meguro-ku, Tokyo 153-8505, Japan

[6] Graduate School of Engineering, Nagoya University, Nagoya 464-8603, Japan

[7] Department of Applied Physics and Physico-Informatics, Keio University, Hiyoshi, Yokohama 223-8522, Japan

*Electronic mail: jun.yoneda@riken.jp

**Recent advances towards spin-based quantum computation have been primarily fuelled by elaborate isolation from noise sources, such as surrounding nuclear spins and spin-electric susceptibility[1-4], to extend spin coherence. In the meanwhile, addressable single-spin and spin-spin manipulations in multiple-qubit systems will necessitate sizable spin-electric coupling[5-7]. Given background charge fluctuation in nanostructures, however, its compatibility with enhanced coherence should be crucially questioned[8-10]. Here we realise a single-electron spin qubit with isotopically-enriched phase coherence time (20 μs)[11,12] and fast electrical control speed (up to 30 MHz) mediated by extrinsic spin-electric coupling. Using rapid spin rotations, we reveal that the free-evolution dephasing is caused by charge (instead of conventional magnetic) noise featured by a 1/$f$ spectrum over seven decades of frequency. The qubit nevertheless exhibits superior performance with single-qubit gate fidelities exceeding 99.9% on average. Our work strongly suggests that designing artificial spin-electric coupling with account taken of charge noise is a promising route to large-scale spin-qubit systems having fault-tolerant controllability.**

Single electron spins confined in quantum dots (QDs) constitute a promising semiconductor platform for spin-based quantum computing. A wide variety of devices[11-20], have been examined in a quest for higher quality factor, $Q$ – the number of qubit operations available before coherence is lost. Fault-tolerant universal gates in this system[21,22] would comprise a sequence of several single-qubit operations and require single-qubit $Q \sim 1000$. $Q$ benefits both from a long dephasing time ($T_2^*$) and a short manipulation time ($T_\pi$), between which a coherence-controllability trade-off has been experimentally identified.

Unlike charge, a single spin does not couple directly to electric noise, so dephasing is predominated by magnetic noise, typically from surrounding nuclear spins. A major approach to improve $Q$ has therefore been engineering host materials to enhance $T_2^*$ by suppressing magnetic fluctuations. With the ultimate development of isotopically purified devices[11,12], $Q$ reaches $\sim 100$, achieved by slow spin manipulation as a manifestation of the coherence-controllability trade-off. Note that now the spin dephasing mechanism becomes controversial, with no longer apparent magnetic fluctuators – but there is no convincing experimental study that elucidates non-magnetic dephasing. On the other hand, an approach to increase $Q$ by shortening $T_\pi$ has been commonly pursued in materials with large intrinsic spin-electric coupling (SEC), e.g. narrow-bandgapped semiconductors[15,16], bent nanotubes[17] and holes in silicon[20]. While this will promote individual accessibility[6,7], realising $Q > 100$ in such structures has also proven challenging due to the trade-off; $T_2^*$ is degraded by isotopically-purified material standards.

Our work strikes a balance between controllability (short $T_\pi$) and coherence (long $T_2^*$) at an isotopically-enhanced level in order to go beyond the above described approaches, and reveals the coherence limited by electrical charge noise. We introduce "artificial" SEC fields with local magnets[5,14] to an isotopically-clean silicon QD qubit, only to the extent that it barely affects $T_2^*$ while reducing $T_\pi$ by two orders of magnitude. Fast spin rotations allow us to unveil that, as a result of inducing SEC in the absence of nuclear spins, the free-evolution dephasing of the spin is caused solely by $1/f$ charge noise at least up to the sub-MHz range. This is in marked contrast to conventional dephasing due to magnetic noise sources featured with higher spectral exponents[12-17]. With the Rabi oscillation $Q$ reaching 888, we demonstrate $> 99.9\%$ average single-qubit control fidelity.

The experiment is performed on an electron spin confined in a $^{28}$Si/SiGe QD by applying electrical pulses to initialise, control and read out the qubit state (Fig. 1a,b and Methods). Our qubit device is equipped with a proximal micro-magnet which induces SEC fields both in the "transverse" and "longitudinal" coupling directions (Fig. 1c). The transverse field slope $b_{trans}$ mediates rapid electrical spin rotations[19,23], while the longitudinal slope $b_{long}$ provides electrical tunability of the qubit frequency, which would be useful for selective manipulation and non-demolition dispersive readout[6,7]. Aside from improving qubit controllability, these SEC fields mediate dephasing in combination with the fluctuating electrical field (charge noise), $\delta E_{rms} \sim 0.1$ V/cm typically[9,10]. In our nuclear-spin-free device, this otherwise negligible effect will become relevant. To still benefit from isotopic purification would require $b_{long} \ll 3$ mT/nm (Supplementary Materials), assuming the orbital energy spacing $\Delta_{orb} \sim 1$ meV. We therefore pattern our Co micro-magnet[24] such that $b_{long} \sim 0.2$ mT/nm, which in practice limits $b_{trans}$ to $\sim 1.0$ mT/nm.

To first demonstrate enhanced transverse SEC, we drive electric dipole spin resonance (EDSR). When the control microwave is exactly on resonance, the spin-up probability $P_{up}$ shows a so-called Rabi oscillation as a function of the microwave duration. Figure 1d shows 16.6 MHz oscillations, whose decaying time is too long to measure within forty π rotations. The decaying time of 3.9 MHz Rabi oscillations is 113±3 μs, yielding the Rabi oscillation $Q = 888\pm25$ (Supplementary Fig. 2b). The rotation frequency increases proportionally to the microwave amplitude up to approximately 20 MHz (Fig. 1e), above which faster oscillation damping is observed (Supplementary Fig. 2a). A nearly-ideal EDSR rotation in the linear regime is further verified by the chevron pattern (Fig. 1f); the pattern reflects the qubit spin rotation along a tilted axis in the Bloch sphere with deliberately detuned microwave excitation.

We next quantify a longitudinal SEC field in the device. This is performed by applying an additional bump pulse to gate R in the control stage, during which the qubit precession frequency is rapidly shifted (Fig. 2a). Figure 2b shows the resulting phase-shift-induced oscillations of $P_{up}$ lasting 20 μs with no indication of decay. The phase rotation speed grows linearly with the bump amplitude $\delta V_R$, yielding a frequency-shift lever-arm of 93 kHz/mV. Note that the maximum shift (5.2 MHz) is limited by the pulse-generating hardware and induced within the same equilibrium charge occupation.

A crucial question, given the enhanced electrical controllability, is whether the isotopically-enhanced spin coherence survives the induced size of SEC. We reveal this by measuring $T_2^*$ from the Ramsey interference effect (Fig. 3a). Curve fitting to the fringe decay yields $T_2^* = 20$ μs, consistent with the EDSR spectral width (Fig. 3b). This $T_2^*$ value is similar to those in $^{28}$Si QDs without an on-chip magnet[11,12,25] and indicates the compatibility of the SEC-enhanced controllability and the isotope-purified coherence. It also suggests that $T_2^* \sim 2$ μs previously reported for isotopically natural Si/SiGe QDs (refs. 18,19) is limited by the fluctuating nuclear-spin field due to 5% $^{29}$Si.

To further study phase coherence of our qubit, we employ Carr-Purcell-Meiboom-Gill (CPMG) protocols[26,27]. Such dynamical decoupling controls can partially cancel the dephasing effect, with efficacy strongly dependent on the qubit noise spectral density, $S(f)$. The CPMG coherence time, $T_2^{\mathrm{CPMG}}$, characterises the decay timescale of the remaining phase coherence $A_{\mathrm{CPMG}}$ (echo amplitude) after the sequence with the total wait time $t_{\mathrm{wait}}$. Note that in the following we normalise $A_{\mathrm{CPMG}}$ by the estimated measurement visibility for each number of π pulses, $n_\pi$. When $n_\pi = 1$, it is essentially a Hahn echo sequence, and the measured coherence time is 99±4 μs (Fig. 3c).

When $n_\pi$ is sequentially increased as $n_\pi = 2^1, 2^2..., 2^{10}$, $A_{\mathrm{CPMG}}$ always takes the form of $\exp[-(t_{\mathrm{wait}}/T_2^{\mathrm{CPMG}})^\alpha]$ (Fig. 4a) with the best fit values of α falling in the range of 2.2±0.2. We see a clear power-law scaling[27] of $T_2^{\mathrm{CPMG}}$ with $n_\pi$, i.e. $T_2^{\mathrm{CPMG}} \propto n_\pi^\beta$ with the exponent β = 0.526±0.011 (Fig. 4a inset). We furthermore relate $A_{\mathrm{CPMG}}$ to $S(f)$ using the filter function formalism for Gaussian noise[26], and obtain (Supplementary Materials) for $n_\pi \gtrsim 8$,

$$S(n_\pi/2t_{\mathrm{wait}}) \simeq -\ln(A_{\mathrm{CPMG}})/2\pi^2 t_{\mathrm{wait}}. \qquad (1)$$

Figure 4b plots $S(f)$ calculated from Eq. (1) for $n_\pi \geq 8$ and $0.15 < A_{\mathrm{CPMG}} < 0.85$. All data points nicely follow a power law $1/f^{\,\gamma}$ in the frequency range of 13 to 320 kHz, with γ = 1.01±0.05. From this simple power-law spectrum, we obtain α = γ+1 = 2.01 and β = γ/(γ+1) = 0.500, both of which agree well with the independent fitting results to the $A_{\mathrm{CPMG}}$ envelopes and the $T_2^{\mathrm{CPMG}}$ scaling described above. Such $1/f$ charge noise has commonly been observed in electrical properties in semiconductor devices[8-10] and is also measured as current fluctuations in our device (Supplementary Fig. 1).

We can independently estimate $S(f)$ around 0.01-1 Hz by tracking Ramsey fringe dynamics[28] (Fig. 4b and Supplementary Materials). Surprisingly enough, the results fall on the extended line of the noise spectrum revealed by CPMG at tens of kHz. This means that the $1/f$ charge noise constitutes the only dominant source of spin phase noise in this device over seven decades of frequency. Indeed, this scenario predicts $T_2^*$ of 25 μs, in excellent agreement with the value measured directly (Supplementary Materials).

As a concluding test of how SEC enhances the qubit performance, we characterise single-qubit control fidelities based on randomized benchmarking[29] (see Methods). As the number of applied Clifford gates $m$ increases, the standard sequence fidelity (Fig. 5) decays as $Vp_C^m$, with the visibility $V = 0.712$ and the depolarising parameter $p_C = 0.99721 \pm 0.00009$. This corresponds to the single Clifford gate fidelity of 99.861±0.005% and the average single gate fidelity of 99.926±0.002%, which give nearly an-order-of-magnitude smaller error rates than the best values reported in QDs[11,19] and well exceed the threshold for fault-tolerant quantum computing[21].

We also evaluate the fidelity of each single-qubit operation individually through the interleaved benchmarking (Methods). The obtained fidelity curves (Fig. 5) are then fitted with $V(p_C p_G)^m$, yielding the gate fidelity as $(1+p_G)/2$. The average value of fidelities 99.928% is consistent with the value from the standard sequence. We speculate that the gate fidelities are limited by systematic pulse calibration errors rather than dephasing effects, since the fidelities for half π rotations are strongly sign dependent.

To conclude, designing SEC in QD spin qubits is a viable approach to meet antithetic requirements for high electrical controllability and isotopically-enriched spin coherence. In this approach, $1/f$ charge noise can become the exclusive source of free-evolution dephasing of a single spin, and the qubit performance may be further enhanced by improving the electrical stability. Qubit control fidelity exceeding 99.9% strongly underpins the prospects for fault-tolerant universal quantum computation in this architecture.

**Acknowledgements**

We thank the Microwave Research Group in Caltech for technical supports. This work was supported financially by CREST, JST (JPMJCR15N2, JPMJCR1675) and the ImPACT Program of Council for Science, Technology and Innovation (Cabinet Office, Government of Japan). JY, TN and TO acknowledge support from RIKEN Incentive Research Projects. TO acknowledges support from PRESTO (JPMJPR16N3); JSPS KAKENHI Grant Numbers JP25800173 and JP16H00817; Strategic Information and Communications R&D Promotion Programme; Yazaki Memorial Foundation for Science and Technology Research Grant; Japan Prize Foundation Research Grant; Advanced Technology Institute Research Grant; the Murata Science Foundation Research Grant; Izumi Science and Technology Foundation Research Grant; TEPCO Memorial Foundation Research Grant; The Thermal & Electric Energy Technology Foundation Research Grant; The Telecommunications Advancement Foundation Research Grant; Futaba Electronics Memorial Foundation Research Grant; and MST Foundation Research Grant. TK acknowledges support from JSPS KAKENHI Grant Numbers JP26709023, JP16F16806. ST acknowledges support by JSPS KAKENHI Grant Numbers JP26220710 and JP16H02204.

**Author contributions**

J.Y. performed the bulk of measurement and data analysis. K.T. fabricated the device with the help of T.O. Y. H., N. U., and K. M. I. supplied the isotopically enriched Si/SiGe heterostructure. J.Y. wrote the manuscript with inputs from other authors. T.N. M.R.D., G.A., T.N., T.K., and S.O. contributed to device fabrication and measurement. S.T. supervised the project.

**Competing financial interests**

The authors declare that they have no competing financial interests.

**Figure 1 | QD device with extrinsic SEC fields.**

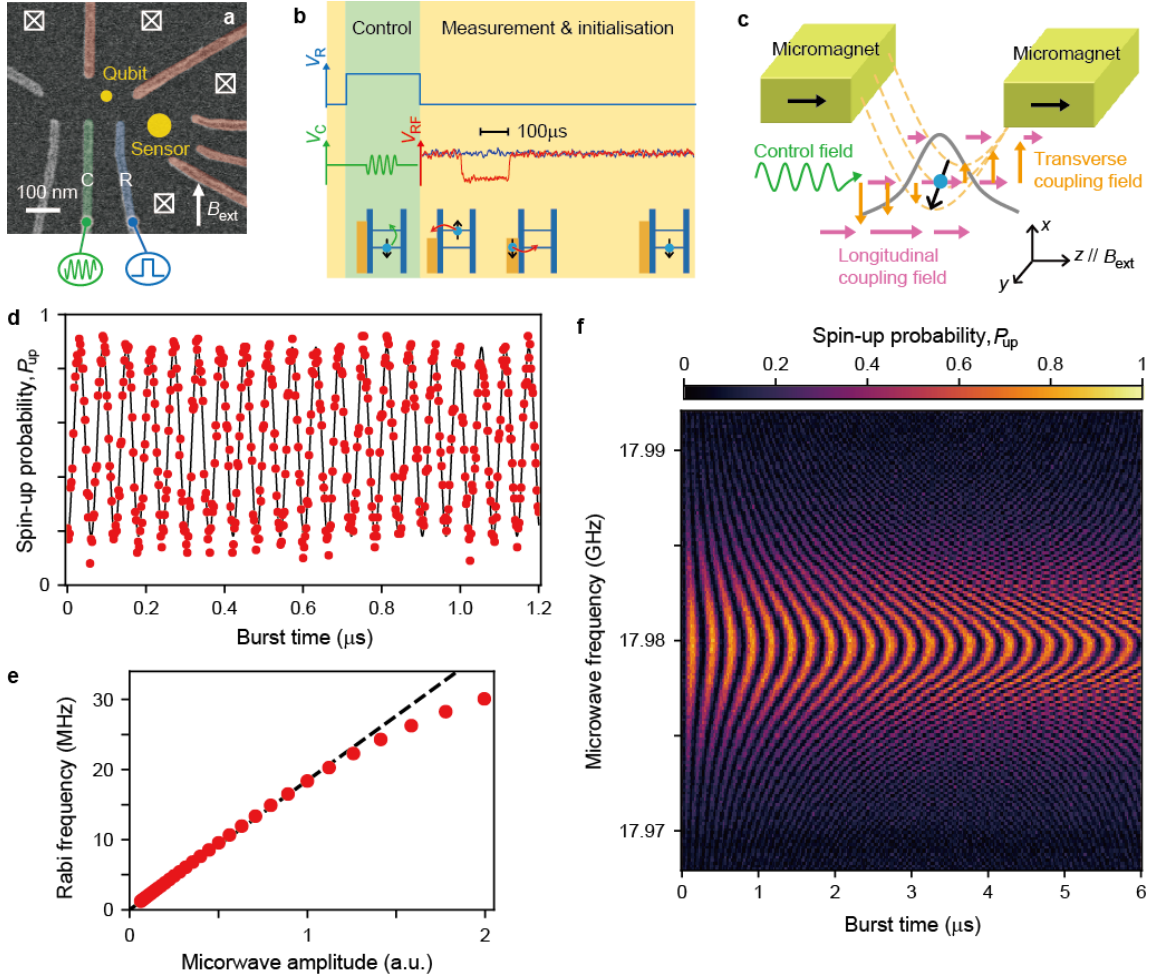

(**a**) Device layout. Circles indicate approximate positions of a qubit and a charge sensing QD. We typically apply an external magnetic field $B_{ext}$ of around 0.5 T so that the Larmor precession frequency is roughly 18 GHz (the optimal frequency for the set-up of our control circuit). (**b**) Control pulse sequence. Waveforms are applied to gate electrode potentials $V_R$ and $V_C$ of gates R and C, respectively. Traces of typical radio-frequency charge-sensing signals $V_{RF}$ with and without tunnelling events are shown in the inset. (**c**) Micro-magnet SEC fields. The magnet is designed to induce a spatially inhomogeneous stray field $B_{MM}$ at the QD position when magnetised along $B_{ext}$. The transverse coupling is produced by the inhomogeneous component perpendicular to $B_{ext}$ and is proportional to the field slope $b_{trans} = (\vec{e} \cdot \nabla) B_{MM}^{x}$, where $\vec{e}$ is the unit vector along an in-plane ($yz$) electric field. The longitudinal one is, on the other hand, mediated by the gradient of the parallel component $b_{long} = (\vec{e} \cdot \nabla) B_{MM}^{z}$. We have assumed a QD confinement that is strong vertically (along $x$) and symmetric laterally. (**d**) Rabi oscillation. Each data point represents the probability of detecting tunnelling events, which we interpret as $P_{up}$, based on 100 single-shot measurements. The solid curve is the best-fit cosine with the Rabi frequency of 16.6 MHz (no decay is assumed). The oscillation visibility is limited by the initialisation/readout fidelity. (**e**) Driving amplitude dependence of the Rabi frequency. The dashed line plots a linear fit with the data points whose Rabi frequencies are below 24 MHz. (**f**) Chevron pattern. $P_{up}$ is collected as a function of microwave burst time and detuning. The Rabi frequency is 3.9 MHz and $B_{ext}$ is 0.506 T.

**Figure 2 | Longitudinal SEC characterisation.**

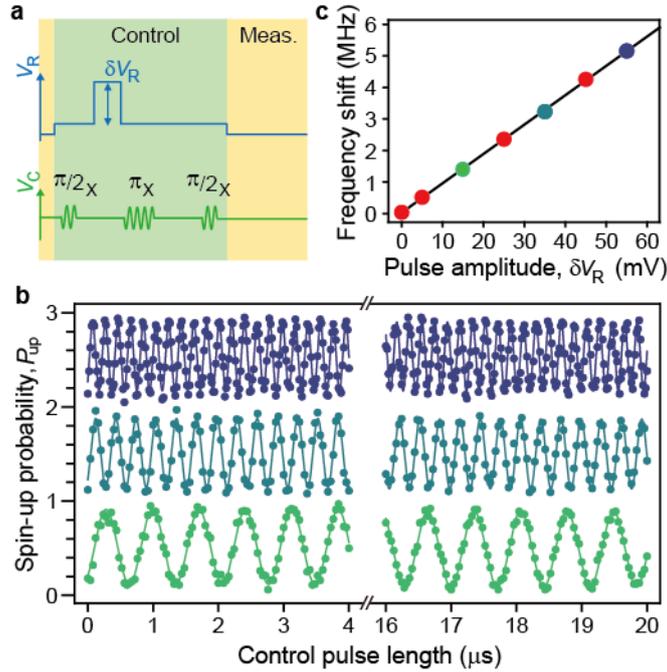

(**a**) Schematic representation of pulse sequences used to evaluate the longitudinal SEC. To detect the pulse-induced phase, we use three equidistant EDSR rotations around a fixed axis (*x*): $\pi/2_X - \pi_X - \pi/2_X$. The first and the last $\pi/2$ rotations together map the phase accumulated between them to a measurable spin component. For a $\pi$ phase shift, for example, an electron initialised in the spin-down state ends up in the spin-up state after the pulse sequence. By inserting a $\pi$-flip midway, the signal becomes robust against a static, microwave detuning effect. (**b**) Oscillation due to phase rotation for pulse amplitudes $\delta V_R$ = 15, 35 and 55 mV. Measurement data with control pulse lengths between 0 and 4 µs or between 16 and 20 µs are shown. Solid curves are the fits with a sinusoidal function with a phase offset (common to both regions). The oscillation frequency gives the size of the induced qubit-frequency shift. The traces are offset by 1 for clarity. (**c**) Pulse amplitude dependence of the extracted frequency shift. The solid line plots a linear fit, giving the gate lever-arm of 93.1 ± 0.7 kHz/mV for gate R. A similar lever-arm is obtained in the frequency domain (Supplementary Fig. 3).

**Figure 3 | Phase coherence and a spin echo.**

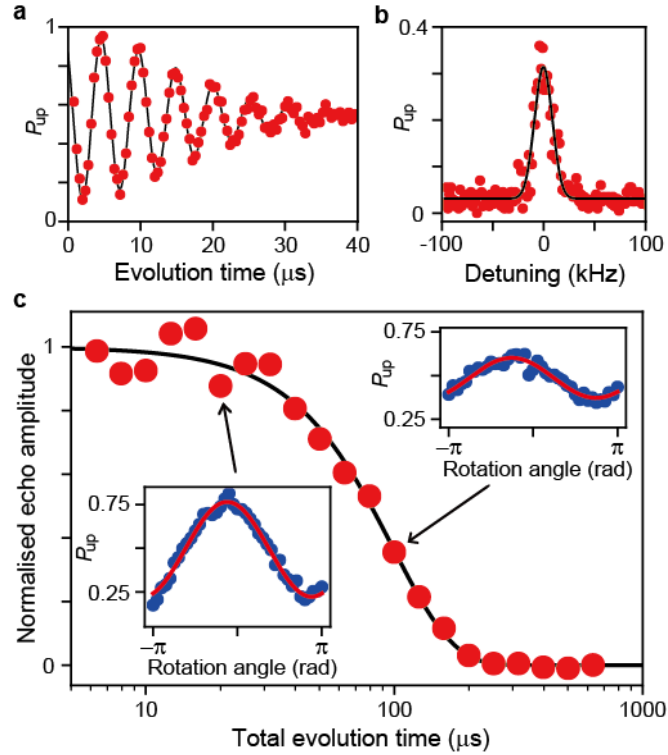

(a) Ramsey interference. The Ramsey fringe is observed for a spin undergoing two EDSR half-π rotations, i.e. π/2$_X$ − π/2$_X$. The visibility reflects the phase coherence preserved during the inter-rotation period. The final $P_{up}$ oscillates at a microwave detuning frequency, with a characteristic damping time giving a measure of $T_2^*$. Each data point is the average over 500 single-shot measurements per evolution time, and the total measurement time is 98 sec. The curve shows the fit with $A\cos(2\pi ft + \varphi)\exp(-(t/T_2^*)^\alpha) + B$ where $t$ denotes the inter-pulse evolution time. The best fit is obtained with $A = 0.43$, $B = 0.54$, $f = 194$ kHz, $T_2^* = 20.4$ μs, and $\alpha = 2.0$. (b) EDSR spectrum at a low excitation power fitted with a Gaussian peak function. The full width at half maximum is 21 kHz. The lateral axis is a microwave frequency offset from 17.980347 GHz. To avoid extrinsic spectral broadening, a three-ms-long, Gaussian-envelope burst is used. Note that the peak height is limited due to the slow Rabi frequency (estimated to be ~ 1 kHz). (c) Normalised echo signal (CPMG with $n_\pi = 1$) as a function of total evolution time, $t$. The sequence comprises three EDSR pulses with different rotation axes, π/2$_X$ − π$_Y$ − π/2. Each data point represents the oscillation amplitude $A_{CPMG}$ of $P_{up}$ by sweeping the rotation axis of the final echoing π/2 rotation (see the insets). Fitting with a form of $\exp(-(t/T_2^{\mathrm{CPMG}})^\alpha)$ yields $T_2^{\mathrm{CPMG}} = 99$ μs and $\alpha = 1.8$.

**Figure 4 | Dynamical decoupling and noise spectral density.**

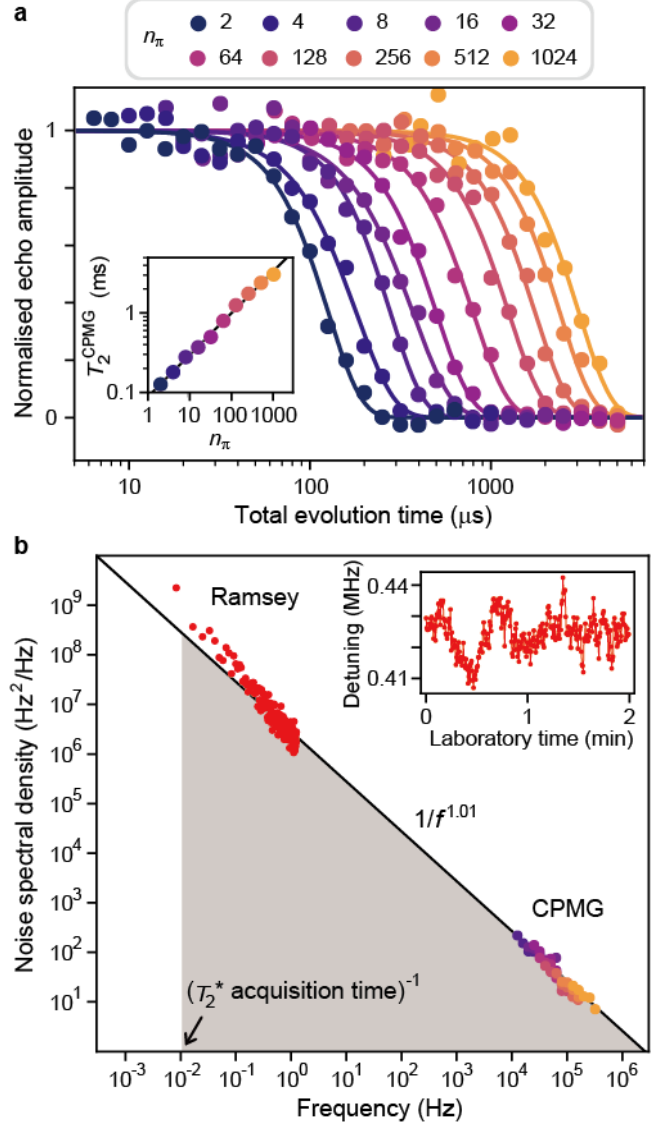

(**a**) Coherence decay under CPMG decoupling pulses as a function of total evolution time, for different $n_\pi$. In the CPMG sequence, equally-spaced π pulses are applied $n_\pi$ times between two π/2 rotations. The rotation axis of the decoupling $\pi_Y$ pulses is set orthogonal to that of the first $\pi/2_X$ pulse. For the largest $n_\pi = 1024$ we obtain $T_2^{\mathrm{CPMG}} = 3.1$ ms, which is two-orders-of-magnitude longer than $T_2^*$. Similarly to Fig. 3(c), each data point represents the coherent oscillation amplitude $A_{\mathrm{CPMG}}$ extracted by changing the angle of the last π/2 rotation. Since waveforms describing EDSR pulse sequences are at least as long as the total evolution time, the sampling time resolution needs to be sufficiently large. Therefore, relatively long rectangular-shaped bursts are used here (500 and 250 ns for π and π/2 rotations, respectively). The inset shows the extracted coherence times $T_2^{CPMG}$ as a function of $n_\pi$. $T_2^{CPMG}$ grows monotonically with $n_\pi$, as expected for coherence limited by low-frequency noise. Fitting is performed for values in the logarithmic scale. (**b**) Noise spectral content extracted independently from $A_{CPMG}$ in (**a**) and from repeated Ramsey measurements. The solid line is a fit of the CPMG noise data, $S \propto 1/f^{1.01}$. Ramsey noise data (red points) are calculated from the estimated qubit frequency evolution of 24 mins in total. They are based on 3600 Ramsey fringe records with each containing 200 single shot measurement results. An example of a two-minute trace of the qubit frequency is displayed in the inset.

**Figure 5| Gate fidelity benchmark.**

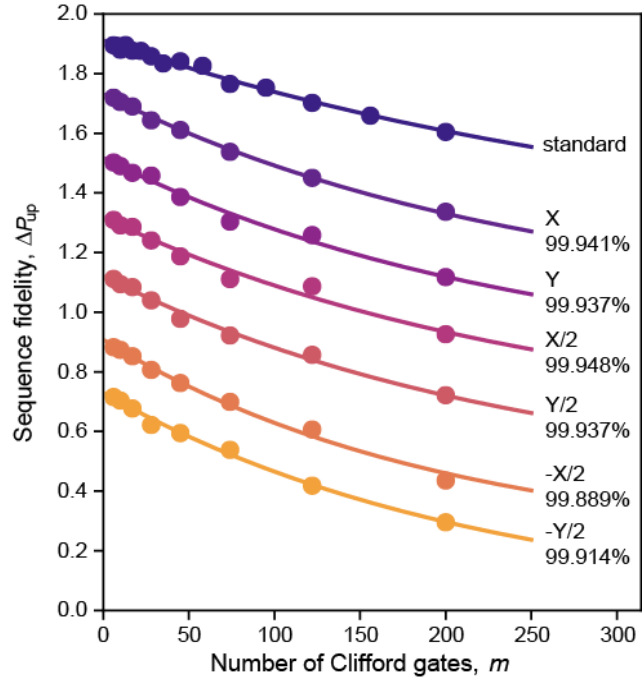

Sequence fidelities for standard (top-most) and interleaved randomised benchmarking show a single-exponential decay over more than 200 Clifford gate operations. Interleaved single-qubit gates are annotated in the figure along with extracted fidelities. Traces are offset by an increment of 0.2 for clarity. Visibilities are all within $0.72 \pm 0.012$.

**Methods**

**The Device.** The Si/SiGe heterostructure was grown by gas-source molecular beam epitaxy. An isotopically purified silane source (with a residual concentration of $^{29}$Si at around 800 ppm) was used exclusively for the strained well. The QD and the bottom of the 250-nm-thick on-chip micro-magnet are separated by ~ 170 nm and designed in line with ref. 24. Carriers are induced in the well with a global top-gate electrode and further lateral confinement is provided by individually biasing surface metal gates. We confirm the valley splitting is much larger than the Zeeman splitting from magneto-spectroscopy measurements.

**Measurement pulses.** We apply two-stage pulses to gate R for spin initialisation, control and readout as shown in Fig. 1b, except for longitudinal SEC characterisation measurements (Fig. 2a). The typical readout time is ~ 1 ms, as we tune the QD-reservoir tunnel rate to around 10 kHz. In the initialisation and readout stage, the centre of the Zeeman-split energy levels in the QD is aligned to the chemical potential in the reservoir. Then, only the excited spin-up electron will tunnel out, and only the ground spin-down electron will tunnel in. We detect such tunnelling events by thresholding charge sensor reflectometry signals, to measure the qubit spin state in a single-shot manner. During the control stage, on the other hand, both spin sublevels in the QD are plunged energetically far below to prevent population leakage. During spin manipulation, the measurement reflectometry carrier power is blanked.

**Clifford benchmarking protocol.** Randomized benchmarking protocols apply sequences of randomly chosen gates from some group and map the error probability to the fidelity decay rate as a function of sequence length. In the standard randomized benchmarking, after initialisation in a spin-down state, we apply $m$ successive Clifford gates, which are randomly chosen to twirl gate errors. In the interleaved benchmarking, we insert an additional test gate between randomly chosen Clifford gates. We then rotate the spin state such that the output ideally becomes one of the target spin states, up or down[29]. We calculate the difference of the spin-up probabilities $\Delta P_{up}$ between the two target-state cases with each based on 1000 single-shot measurements. Each sequence fidelity value is obtained by averaging $\Delta P_{up}$ over 20 different random Clifford gate sets. The sequence fidelity then decays as $(2F - 1)^m$ with $F$ denoting the average fidelity per step.

      Each of the 24 gates in the Clifford gate set is constructed by concatenating on

average 1.875 single-qubit physical gates of 7 kinds, π and ± π/2 rotations around the *x* and *y* axes and an identity operation. To implement the physical gates, we use two-quadrature EDSR microwave pulses. On the main quadrature, we feed Gaussian envelope bursts truncated at ± 2σ (σ denotes the standard deviation). On the second quadrature, we simultaneously apply the derivative-of-Gaussian envelope pulses to account for the control-induced frequency shift. These pulse amplitudes are individually optimized for π and π/2 gates. The pulse durations (4σ) are 120 and 60 ns for the π and π/2 gates, respectively, due to pulse-shaping hardware limitation.

# Supplementary Information:
# A >99.9%-fidelity quantum-dot spin qubit with coherence limited by charge noise

1. Details of the experimental setup

The device was cooled in a dilution refrigerator with an electron temperature of around 100 mK (estimated from the charging line width). The high-frequency electrical cable connected to gate R was coupled to a DC bias through resistive dividers at room temperature and was then filtered inside the refrigerator with a cut-off frequency of ~ 100 MHz. Gate pulse signals were generated by Tektronix AWG520. The other high-frequency cable for gate C was designed for a bandwidth of 20 GHz. For microwave pulse shaping, we used either analogue (pulse and phase) modulations (for higher powers) or digital, quadrature modulation (for narrower spectral broadening). These modulations were realised by feeding two-channel waveform signals from Tektronix AWG7122C, triggered by Tektronix AWG520 during the control stage. The microwave was generated by Agilent generators E8257D or E8267D (depending on pulse shaping schemes).

We formed a tank circuit resonating at 185 MHz on the device printed circuit board. Reflected RF charge-sensing signals were collected with AlazarTech digitizer ATS9440 at 1 MS/s after demodulation. Single-shot spin measurements were performed by thresholding the maximum signal level with respect to the median value during the measurement stage. This readout thresholding value and the readout gate voltage were calibrated almost hourly prior to measurements such that the Rabi oscillation visibility would be maximised. The microwave carrier frequency was also calibrated similarly often based on Ramsey fringes to compensate for a systematic decrease of the EDSR frequency. We believe that this drift is primarily due to the field decay of the superconducting magnet operated in persistent mode (while also sending a current through the external leads).

2. Charge-noise-induced spin dephasing mediated by a longitudinal SEC

$1/f$ charge noise is commonly observed in nano-electronic devices, and could arise from a variety of mechanisms[8-10]. For QDs based on semiconductor heterostructures, charge traps near interfaces are typically believed to be the leading source. Exposure to SEC will couple the qubit spin to such electric field noise, to which the pure electron spin would be insensitive. More specifically, fluctuating in-plane electric field $\delta E_{\text{rms}}$ will displace the QD position by

$$\delta r_{\text{rms}} = \frac{h^2 e}{4\pi^2 m_{\text{Si}} \Delta_{\text{orb}}^2} \delta E_{\text{rms}},$$

where $h$ is Planck's constant, $e$ is the single electron charge, and $m_{\text{Si}}$ is the electron effective mass in Si (~ 0.2 $m_0$ with $m_0$ the electron rest mass). We have assumed an in-plane symmetric harmonic confinement potential, characterised by the orbital spacing $\Delta_{\text{orb}}$. In electrically controlled QDs, the size of fluctuating electric potential $\delta V_{\text{rms}}$ has been conventionally measured instead of $\delta E_{\text{rms}}$, with a typical value of ~ 1 μeV[8-10]. Given that this level of potential noise is generated by an ensemble of interface charge traps located typically $d_{\text{trap}}$ ~ 100 nm away from the QD[9],

$$\delta E_{\text{rms}} \sim \delta V_{rms}/(ed_{\text{trap}}) \sim 0.1 \text{ V/cm}.$$

This corresponds to $\delta r_{\text{rms}} = 4$ pm for $\Delta_{\text{orb}} = 1$ meV. The QD displacement will shift the qubit frequency via a longitudinal SEC field by

$$\delta f_E = g\mu_B b_{\text{long}} \delta r_{\text{rms}}/h,$$

where $g$ ~ 2 is the Landé $g$-factor in Si. We note that since $B_{\text{ext}}$ is much larger than the size of stray field, the dephasing will be nominally mediated by the longitudinal one. To benefit from isotopic purification, this charge-noise-induced detuning fluctuation $\delta f_E$ should be safely smaller than the nuclear-spin-induced detuning fluctuation $\delta f_{\text{nat}}$ ~ 0.3 MHz (r.m.s.) observed in isotopically natural Si/SiGe QDs[18]. We will then arrive at

$$b_{\text{long}} \ll \frac{4\pi^2 m_{\text{Si}} \Delta_{\text{orb}}^2}{g\mu_B eh\, \delta E_{\text{rms}}} \delta f_{\text{nat}} \sim 3 \text{ mT/nm}.$$

3. Charge noise spectrum in the device

We evaluate the spectrum of the device charge noise by recording the demodulated RF amplitude of the sensor signal for 0.5 sec. We divide the trace into 25 parts (the Bartlett method) and calculate the average power spectral density (Figure S1). The procedure is repeated for cases where the sensor sensitivity is maximal and minimal (tuned to a Coulomb slope and valley). The obtained spectra in a frequency range of 10 – 1000 Hz show clear evidence of 1/$f$ charge noise in the device.

4. Rabi oscillation decay at moderate and strongest driving powers

The Rabi oscillation $Q$-factor is a key parameter to quantify the spin-qubit controllability and can be defined as $Q = T_2^{Rabi}/T_\pi = 2\,T_2^{Rabi} f_{Rabi}$, where $T_2^{Rabi}$ denotes the characteristic decaying time of the Rabi oscillation and $1/T_\pi = 2f_{Rabi}$ is the $\pi$ rotation time. While both $T_2^{Rabi}$ and $f_{Rabi}$ should ideally increase as we raise the microwave amplitude, it is evident from Figure 1e of the main text that $f_{Rabi}$ does not linearly grow above 20 MHz in the present device. Furthermore, when the spin is driven in this saturating range, we observe faster Rabi oscillation decay (Figure S2a). These observations imply that the Rabi oscillation $Q$-factor would be peaked at some $f_{Rabi}$ value, which would be an educated estimate of an optimal working point to enhance the control fidelity[19]. In our current setup for randomised benchmarking experiments, however, the maximum applicable amplitude is decreased by the conversion loss of an additional mixer for quadrature control (Marki Microwave MLIQ-0218), making maximum $f_{Rabi} \sim 4$ MHz, far below saturation. The Rabi oscillation quality factor for $f_{Rabi} = 3.93$ MHz is evaluated to be $Q = 888$ (Figure S2b).

5. Frequency-domain measurement of in-situ qubit-frequency shift

In this section we describe a measurement of the qubit-frequency shift in the spectral domain, as opposed to the one in the time domain (Figure 2 in the main text). We obtain the resonance frequency $f_{res}$ at different operation biases by changing the control pulse amplitude $\Delta V_R$ (see Figure S3a). We measure frequency-dependence of $P_{up}$ under a $\pi$ pulse (we fix the Rabi frequency $f_{Rabi}$ to 1 MHz and the microwave duration to 500 ns) and fit the obtained spectra to a model function for an ideal $\pi$ rotation,
$$A\,|\pi \sinh(D(f))\,/2D(f)|^2 + B$$
where $D(f) = \frac{\pi}{2}\sqrt{i[(f-f_{res})/f_{Rabi}]^2 - 1}$, and $A$ and $B$ account for initialisation and readout errors (Figure S3b). Figure S3c plots the extracted $f_{res}$ values as a function of $\Delta V_R$, from which we get a gate lever-arm of 91.1 kHz/mV for gate R. While, strictly speaking, this measures the steady-state frequency shift, a consistent value is obtained for the real-time modulation as well (see the main text).

6. Calculations of noise spectral content

The noise spectral content $S(f)$ used in the main text is defined by the power spectral density of detuning frequency noise $\nu(t)$, or
$$S(f) = \lim_{T\to\infty} \frac{1}{T} \left| \int_0^T dt\, e^{2\pi i f t}\, \nu(t) \right|^2. \tag{S1}$$
From the Wiener-Khinchin theorem, we can rewrite this as

$$S(f) = \int_{-\infty}^{\infty} dt \ e^{2\pi i f t} \langle v(t')v(t'+t) \rangle$$

in terms of the autocovariance function $\langle v(t')v(t'+t) \rangle = \lim_{T \to \infty} \frac{1}{T} \int_0^T dt' v(t')v(t'+t)$, with $\langle \cdot \rangle$ denoting the statistical average. If the distribution of $v(t')$ is Gaussian, random phase accumulation $\phi(t)$ after some sequence with the total evolution time $t$, $\phi(t) = \int_{-\infty}^{\infty} dt' \ 2\pi \ v(t') \ \theta_t(t')$, is also Gaussian distributed. Note we have introduced a pulse-sequence specific function $\theta_t(t')$, which switches between $\pm 1$ at every decoupling $\pi$ pulse and is 0 before and after the sequence[26]. Then the coherence amplitude is given by

$$\exp[\langle i\phi(t) \rangle] = \exp\left[-\frac{\langle \phi(t)^2 \rangle}{2}\right] = \exp\left[-4\pi^2 \int_0^{\infty} df \ S(f) \ F_t(f)\right] \quad (S2)$$

where $F_t(f) = \left| \int_{-\infty}^{\infty} df \ e^{-2\pi i f t'} \ \theta_t(t') \right|^2$.

For decoupling pulses, $F_t(f)$ effectively behaves as a band-passing filter. By denoting times at which the decoupling $\pi$ pulses are applied as $\tau_k$ (and with $\tau_0 = 0$ and $\tau_{n_\pi+1} = t$), for even numbered CPMG we have[26]

$$F_{n_\pi, t}(f) = \frac{1}{(2\pi f)^2} \left| \sum_{k=0}^{n_\pi} (-1)^k \left( e^{-2\pi i f \tau_{k+1}} - e^{-2\pi i f \tau_k} \right) \right|^2$$

$$= \frac{4}{\pi^2 f^2} \frac{\sin^4\left(\frac{\pi f}{2 f_{\rm rep}}\right) \sin^2\left(\frac{\pi n_\pi f}{f_{\rm rep}}\right)}{\cos^2\left(\frac{\pi f}{f_{\rm rep}}\right)},$$

where $f_{\rm rep} = n_\pi/t$ is the decoupling pulse frequency. For $n_\pi > 8$, it narrowly peaks at odd multiples of $f_{\rm rep}/2$ with contributions from higher harmonics decaying quadratically. Noting from Parseval's theorem that

$$\int_0^{\infty} df \ F_t(f) = \frac{1}{2} \int_{-\infty}^{\infty} dt' |\tilde{\theta}_t(f)|^2 = \frac{t}{2},$$

we can derive Eq. (1) of the main text from Eq. (S2);

$$\ln(\exp[\langle i\phi(t) \rangle]) = -4\pi^2 \int_0^{\infty} df \ S(f) \ F_t(f) \simeq -2\pi^2 t \ S(f_{\rm rep}/2).$$

For Ramsey experiments ($n_\pi = 0$), $\lim_{ft \to 0} F_t(f) = t^2$. Then for $1/f$ noise $S(f) = S_0/f$, Eq.(S2) reduces to

$$\exp(-(t/T_2^*)^2) \simeq \exp\left[-4\pi^2 t^2 \int_{1/T}^{1/t} df\, S(f)\right] = \exp\left[-4\pi^2 t^2 S_0 \ln\left(\frac{T}{t}\right)\right]$$

which gives

$$1/T_2^* \simeq 2\pi\sqrt{S_0 \ln(T/t)}\,. \tag{S3}$$

Here $t$ is the evolution time and $T$ is given by the total measurement time, typically ~ 100 sec in our case (Fig. 3a of the main text). Then the logarithmic factor is ~ 15 when $t$ is tens of microseconds. Assuming $S_0 = 2.67 \times 10^6$ as extracted from CPMG coherence amplitudes, we get $T_2^* = 24.8$ μs for $t = 20$ μs, in excellent agreement with the value obtained independently from Ramsey fringes.

The discrete version of Eq. (S1) provides another way of estimating $S(f)$ based on Ramsey sequences, which we employ in Figure 4c of the main text. Supposing we have a sequence of $v(t_n)$ of a length $N$ with a sampling lag of $t_d$, the discrete version of $S(f)$ is

$$S_d(f_k) = \frac{t_d}{N}\left|\sum_{n=1}^{N} v(t_n)\, e^{-2\pi i f_k t_d}\right|^2$$

where $f_k = k/Nt_d$. It is known that the average of $S_d(f_k)$ for $M$ sets of such sequences will asymptotically tend to $S(f)$. To perform this spectral estimation, we infer detuning values from Ramsey fringes based on 200 single shot measurements in total (corresponding to a sampling lag of 0.4 seconds) using a Bayesian estimation method[28]. We use evolution times between 0.4 and 40 μs and repeat the cycle 3600 times continuously with $(M, N) = (12, 300)$.

**Figure S1 | Charge noise spectrum in the device.**

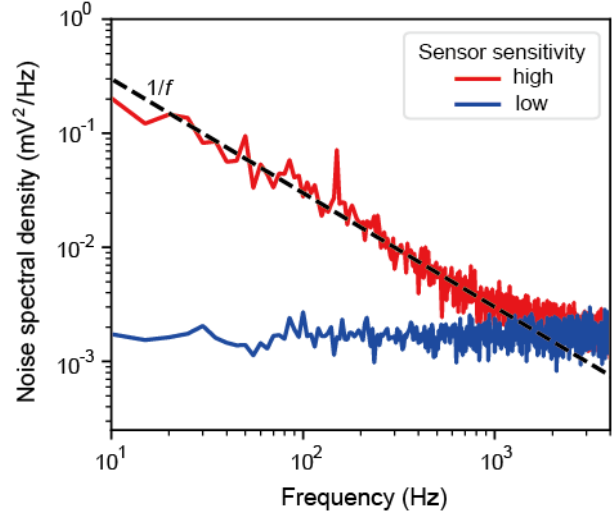

The power spectral density of the device charge noise is estimated by performing discrete Fourier transformation on 25 traces of the sensor signal when the sensor is tuned to a sensing condition (red trace). It shows a clear 1/$f$ dependence below 1 kHz, above which it is exceeded by the background noise floor (blue trace).

**Figure S2 | Rabi oscillation decay.**

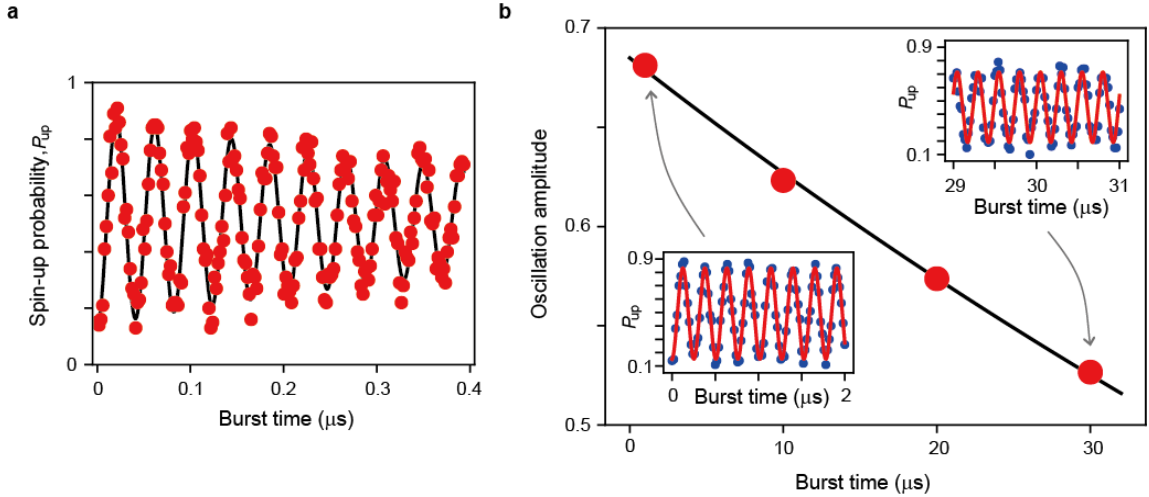

(**a**) Rabi oscillation in the saturating, high-power range. In contrast to the one below 20 MHz (see Fig. 1d of the main text), a clear oscillation decay is observed. The solid curve plots the best fit to $A\cos(2\pi f_{Rabi} t)\exp(-t/T_2^{Rabi}) + B$ with $f_{Rabi}$ = 24.3 MHz, $T_2^{Rabi}$ = 0.57 μs, $A$ = 0.374, $B$ = 0.51, which corresponds to $Q$ = 28. (**b**) Rabi oscillation decay for $f_{Rabi}$ = 3.93 MHz. Each data point represents the Rabi oscillation amplitude for a given burst time (see the insets). Fitting to a single-exponential decay (the black curve) gives $T_2^{\text{Rabi}} = 113 \pm 3$ μs.

**Figure S3 | Spectroscopy of the in-situ qubit-frequency shift.**

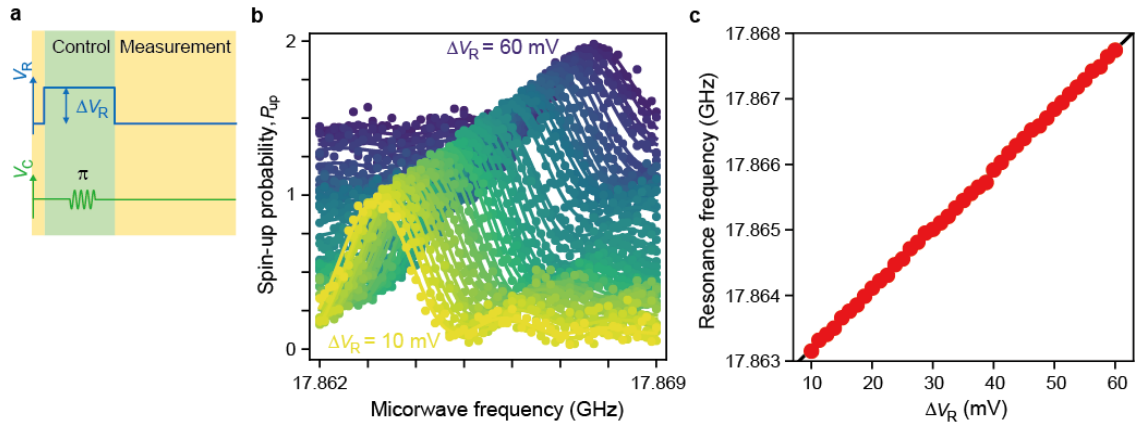

(**a**) Pulse schematic used for spectroscopic measurement of the longitudinal SEC. (**b**) EDSR spectra after a single π pulse for various the control-stage voltages $\Delta V_R$. The traces are offset for clarity. (**c**) Extracted resonance frequency against the control-stage voltage $\Delta V_R$. The linear fit yields the gate lever-arm for the qubit-frequency shift of 91.1 kHz/mV.